\documentclass[%
 reprint,
superscriptaddress,
 amsmath,amssymb,
 aps,
pra,
twocolumn
]{revtex4-2}

\usepackage{graphicx}
\usepackage{dcolumn}
\usepackage{bm}
\usepackage{amsfonts}
\usepackage{physics}
\usepackage[usenames,dvipsnames]{color}

\DeclareUnicodeCharacter{0308}{HERE!HERE!}

\begin{document}

\title{On-demand heralded MIR single-photon source using a cascaded quantum system}

\author{Jake Iles-Smith}
\thanks{These two authors contributed equally}
 \affiliation{Department of Physics and Astronomy, The University of Manchester, Oxford Road, Manchester M13 9PL, United Kingdom}

\author{Mark Kamper Svendsen}
\thanks{These two authors contributed equally}
\affiliation{Max Planck Institute for the Structure and Dynamics of Matter and
Center for Free-Electron Laser Science \& Department of Physics,
Luruper Chaussee 149, 22761 Hamburg, Germany}


\author{Angel Rubio}
\affiliation{Max Planck Institute for the Structure and Dynamics of Matter and
Center for Free-Electron Laser Science \& Department of Physics,
Luruper Chaussee 149, 22761 Hamburg, Germany}
\affiliation{Center for Computational Quantum Physics, Flatiron Institute, New York, New York 10010, USA}
\affiliation{Nano-Bio Spectroscopy Group and European Theoretical Spectroscopy Facility (ETSF), Universidad del Pa\'is Vasco (UPV/EHU), Av. Tolosa 72, 20018 San Sebastian, Spain}

\author{Martijn Wubs}
\affiliation{Department of electrical and photonics engineering, Technical University of Denmark, 2800 Kgs. Lyngby, Denmark
}
\affiliation{
Center for Nanostructured Graphene, Technical University of Denmark, 2800 Kgs. Lyngby, Denmark
}
\affiliation{
NanoPhoton -- Center for Nanophotonics, Technical University of Denmark, 2800 Kgs. Lyngby, Denmark
}

\author{Nicolas Stenger}
\affiliation{Department of electrical and photonics engineering, Technical University of Denmark, 2800 Kgs. Lyngby, Denmark
}
\affiliation{
Center for Nanostructured Graphene, Technical University of Denmark, 2800 Kgs. Lyngby, Denmark
}
\affiliation{
NanoPhoton -- Center for Nanophotonics, Technical University of Denmark, 2800 Kgs. Lyngby, Denmark
}

\date{\today}

\begin{abstract}
We propose a novel mechanism for generating single photons in the mid-Infrared (MIR) using a solid-state or molecular quantum emitter. 
The scheme utilises cavity QED effects to selectively enhance a Frank-Condon transition, deterministically preparing a single Fock state of a polar phonon mode.
By coupling the phonon mode to an antenna, the resulting excitation is then radiated to the far field as a single photon with a frequency matching the phonon mode.
By combining macroscopic QED calculations with methods from open quantum system theory, we show that optimal parameters to generate these MIR photons occur for modest light-matter coupling strengths, which are achievable with state-of-the-art technologies. 
Combined, the cascaded system we propose provides a new quasi-deterministic source of heralded single photons in a regime of the electromagnetic spectrum where this previously was not possible.  
\end{abstract}    

\maketitle

\section{Introduction}
Single photons lie at the heart of many applications in quantum science and technology. Among these applications, quantum metrology and precision spectroscopy stand out as particularly promising, since they can in principle reach the so-called quantum limit of precision 
in transmission and absorption spectroscopy~\cite{Jakeman86,Adesso09,Whittaker17,Moreau17}. 
This approach holds particular promise in the context of spectroscopy of biological systems~\cite{taylor13,wolfgramm13,Taylor15,Samantaray17}, where stringent power limitations are imposed on conventional spectroscopic techniques to prevent damage of biological samples.

Importantly, the range of degrees of freedom that can be explored through single-photon spectroscopy is intricately tied to the frequency of the generated photons. Specifically, single photons in the visible spectrum are instrumental in investigating electronic processes, whereas photons in the mid- and far-infrared (IR) ranges are indispensable for probing vibrational transitions.
Novel quantum light sources in the MIR could enable precise measurements \textit{in vivo} on the single molecular vibration level, understand the fundamental role played by vibrations in quantum matter~\cite{basov2017towards}, as well as give us the spectroscopic means to follow chemical reactions in solvents~\cite{heyne2019infrared,stensitzki2018acceleration}.
%
While there have been significant developments in pushing single-photon sources to the near IR 
and telecommunication C-bands~\cite{vajner24,Phillips24}, these quantum emitters (QEs) rely 
on well defined electronic transitions, 
and thus are not able to access the spectral regimes relevant to vibrational transitions. There are, so far, no known materials with band gaps in the MIR able to efficiently produce single photons with high brightness.

\begin{figure*}
    \includegraphics[width=\textwidth]{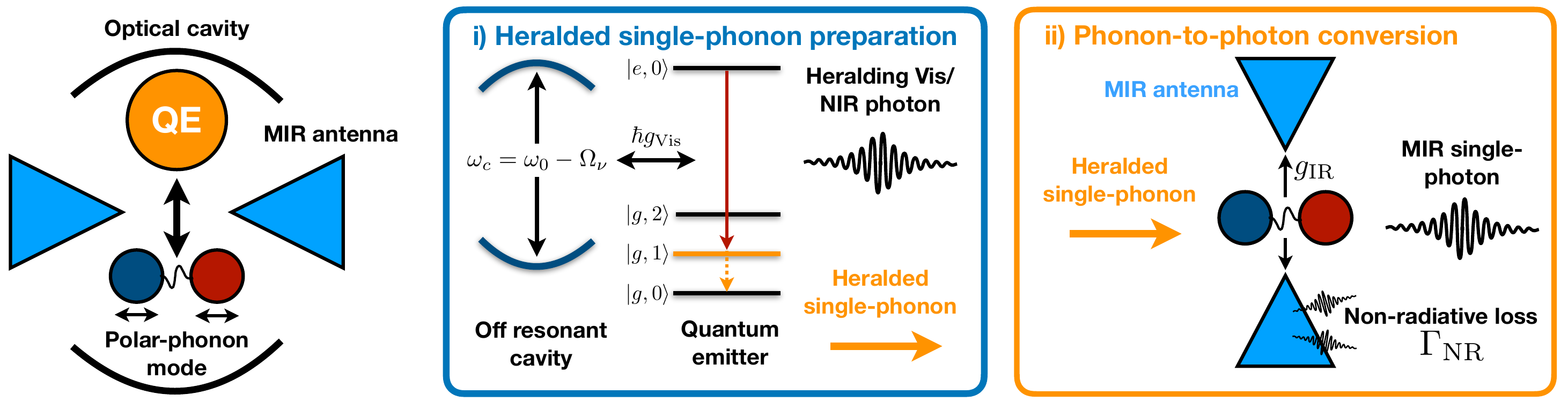}
    \caption{\textbf{A schematic of the MIR-photon generation protocol.} (Left) shows the schematic of a quantum emitter coupled to an optical cavity, interacting with a polar phonon mode. The phonon mode subsequently interacts with a plasmonic or dielectric antenna tuned to the resonance of the phonon mode. The middle and right panels show the two-step photon generation protocol: (i) an initially excited quantum emitter emits a photon in the visible via the first phonon sideband, which is Purcell enhanced through the cavity interaction.
    The emitted photon is used to herald the generation of a single phonon.
    (ii) The phonon then radiates as a single MIR photon through the interaction with the MIR antenna structure.}
    \label{fig:protocol}
\end{figure*}

In this work, we propose a novel scheme for generating single photons in the mid-infrared (MIR) 
using a quantum emitter interacting with polar phonon modes, an {
optical cavity, and a MIR antenna.} 
The protocol consists of a two-part cascaded processes which is shown schematically in Fig.~\ref{fig:protocol}: In the first stage, 
a single Fock state ($\ket{1}$) is prepared in the polar phonon mode through the 
selective Purcell enhancement of the first phonon sideband. As shown schematically in Fig.~\ref{fig:protocol}(i), 
the successful preparation of the single phonon state is heralded by the emission of an optical photon.  
The preparation and manipulation of such quantum vibrational states is of significant interest 
to applications in quantum optomechanics and transduction. 
However, existing proposals are inherently probabilistic~\cite{galland2014prl, riedinger2016non, hong2017sci} 
and demand advanced time-resolved spectroscopic schemes~\cite{velez2019prep}.
In contrast, our protocol can be implemented deterministically with standard 
photonic/plasmonic cavities if the sideband corresponding to the relevant phonon mode 
can be spectrally isolated, and crucially is heralded by the emission of a photon via the cavity mode.

In the second stage of our protocol, the excitation in the phonon mode is radiated as a 
photon by coupling the phonon mode to a resonant antenna structure~\cite{Kats:11,chikkaraddy2023single,roelli2020molecular}, as shown in Fig.~\ref{fig:protocol}(ii). 
By starting with a macroscopic QED description of the quantum emitter and phonon environment, 
we show that the macroscopic dipole moment of the polar phonon mode enables efficient 
transfer of excitation to the antenna structure, 
which is subsequently radiated as a single photon 
with the same frequency as the phonon mode.
Such antennas can be constructed from plasmonic nanostructures
~\cite{Novotny2011} in both the visible and IR~\cite{Biagioni_2012}, and 
allow for the bright emission of single-photons into the far-field~\cite{Koenderink17,hugall18}. 
A similar coupling has been proposed by Weatherly \emph{et al}~\cite{weatherly23} as a means 
to down-convert classical driving fields from visible to IR frequencies. 
In contrast to a recent proposal to generate single photons in the MIR~\cite{groiseau2023single}, our proposed scheme does not depend on a continuous external field driving the emitter in the strong coupling regime, and the emission of the MIR photon is heralded by the presence of a single photon in the visible. Furthermore, our scheme is not restricted to a particular material system, and is applicable to any quantum emitter, solid-state or molecular, coupled to polar phonons.

The manuscript is organised as follows: 
In section~\ref{sec:MIRmodel} we derive a Hamiltonian for the 
emitter-phonon-antenna system that underpins our protocol using 
macroscopic QED. 
In section~\ref{sec:cascaded} 
we characterise the performance of the two-step cascaded system, including generation efficiency and purity of the resulting emitted photon in the MIR. In section~\ref{sec:materials}, we conclude the paper by discussing how the protocol can be realised in material based systems, including defects in two-dimensional (2D) materials, inorganic nanocrystals, and molecular systems.

\section{Model for a cascaded quantum emitter in the Mid IR}
\label{sec:MIRmodel}
We start by considering a molecular system,  with electronic and nuclear degrees of freedom governed by the Hamiltonian $H_\mathrm{mol}$.  Within the length gauge, the interaction between the molecule and the electromagnetic degrees of freedom are given by~\cite{faisal1987theory,tokatly2013time},
\begin{align}\label{eq:LWL_PF_PZW}
   \hat{H} =&  \hat{H}_\mathrm{mol} + \frac{1}{2}\sum^N_{\alpha = 1} \left(\hat{p}_\alpha^2 + \omega_\alpha^2\left[\hat{q}_\alpha + \frac{\boldsymbol{\lambda}_\alpha}{\omega_\alpha}\cdot\boldsymbol{\hat{\mu}}_\mathrm{total}\right]^2\right),
\end{align}
where $\omega_\alpha$, $\hat{p}_\alpha$, $\hat{q}_\alpha$ are respectively the frequency, 
momentum- and displacement operators associated with photon mode $\alpha$.
The total dipole moment can be decomposed into a sum of the nuclear- and electronic dipole moments, 
$ \boldsymbol{\hat{\mu}}_\mathrm{total}  = \boldsymbol{\hat{\mu}}_\mathrm{nu} +\boldsymbol{\hat{\mu}}_\mathrm{el}$. 
The coupling between the molecular degrees of freedom and the electromagnetic field occurs through the matrix 
elements $\boldsymbol{\lambda}_\alpha$. These matrix elements can be related to the dyadic 
Green's function of the electromagnetic environment in a quantitative way and generally reflect the dipole spectral density of the electromagnetic environment~\cite{svendsen2023molecules}. 
We discuss this relation and its implication for realistic structures in the 
supplementary information. 

As our model of the intra-molecular Hamiltonian,  we consider a single polar phonon mode coupled to a two-level molecule,  with ground and excited states,  $\ket{g}$ and $\ket{e}$ respectively.
In the diabatic basis,  and assuming linear electron-phonon coupling, the molecular Hamiltonian takes the form~\cite{weatherly23},
\begin{equation}
   \hat{H}_\mathrm{mol}= (E_\mathrm{vert} - \Omega_\nu Q_e^{(0)}\hat{Q}_\nu)|e\rangle\langle e| + \frac{1}{2}\left(\hat{P}_\nu^2 + \Omega_\nu^2\hat{Q}_\nu^2\right).
\end{equation}
Here $\Omega_\nu$ is the frequency of the polar phonon mode, $\hat{Q}_\nu$ ($\hat{P}_\nu$) is the phonon configurational coordinate displacement 
(momentum) operator. $Q_e^{(0)}$ is the configurational coordinate displacement along that phonon mode associated with the minimum of the excited state potential energy surface~(PES) 
relative to that of the ground state.  The transition energy associated with the ground and excited state is given by 
$E_\mathrm{vert} = \hbar \omega_\mathrm{eg} + S\hbar\Omega_\nu$, 
where $S=(Q_e^{(0)})^2/(2\hbar\Omega_\nu)$ is the Huang-Rhys parameter, and $\hbar \omega_\mathrm{eg}$ 
is the energy difference between the minima of the ground- and excited state PES. 
We can express the molecular Hamiltonian within second quantisation notation, 
using $\hat{Q}_\nu = \sqrt{\hbar/2\Omega_\nu}(\hat{b}^\dagger + \hat{b})$, such that
\begin{equation}
 H_\mathrm{mol} = (E_\mathrm{vert} - \hbar\eta_\nu (\hat{b}^\dagger + \hat{b}))\dyad{e}{e} 
 +\hbar\Omega_\nu \hat{b}^\dagger\hat{b},     
\end{equation}
where we have defined $\hbar\eta_\nu = Q^{(0)}_e \sqrt{\hbar\Omega_\nu/2}$.

\subsection{Coupling the electron-phonon system to the electromagnetic environment}

{
In order to study the impact of the electromagnetic environment on the coupled electron-nuclear system, 
we can expand the dipole operator in the diabatic basis. 
The electronic dipole operator takes the standard form $\boldsymbol{\hat{\mu}}_\mathrm{el} = \boldsymbol{{\mu}}_{eg}\hat{\sigma}_x$,  where 
we have defined the Pauli operator $\hat{\sigma}_x = (\dyad{e}{g} + \mathrm{h.c.})$ 
and neglected the permanent dipole contribution.
The nuclear dipole moment, up to first order in the phonon displacement 
and neglecting the permanent dipole moment of the atomic lattice, is, 
$
    \boldsymbol{\hat{\mu}}_\mathrm{nu}(\hat{Q}_\nu) \approx \boldsymbol{\mu}_{\mathrm{N}}\sqrt{\frac{2\Omega_\nu}{\hbar}}(\hat{Q}_\nu - Q_g^{(0)}),
$
where $\boldsymbol{\mu}_\mathrm{N}$ describes the magnitude of the nuclear dipole moment, 
and may be related to the Born charges of the system as will be discussed in Sec.~\ref{sec:materials}.
The above operator describes the change of the dipole moment due to the displacement along 
the phonon mode, which leads to infrared activity. 
}

With an expression for the dipole moment operator in hand, 
we can consider the coupling to the electromagnetic environment 
which we assume consists of two parts, as shown schematically in Fig.~\ref{fig:protocol}: a single-mode optical cavity in the visible range 
with mode frequency $\omega_\mathrm{Vis}$, and a MIR antenna with frequency $\omega_\mathrm{MIR}$.  
As shown by Katz \emph{et al.}~\cite{Kats:11}, 
it is often well justified to approximate the 
antenna structure by a single damped mode.
Both single-mode electromagnetic environments 
are defined by the displacement (momentum) operators 
$\hat{q}_{\alpha}$ ($\hat{p}_\alpha$) with $\alpha = \mathrm{Vis}, ~\mathrm{MIR}$.
Using the Hamiltonian Eq.~\ref{eq:LWL_PF_PZW}, and expanding the dipole moment operator, we have,
\begin{align}\label{eq:PF_LW_full-expansion}
    &\hat{H} = \hat{H}_\mathrm{mol}
    + \sum_{\alpha = \mathrm{Vis},~\mathrm{IR}}\frac{1}{2}\left(\hat{p}_\alpha^2 + \omega_\alpha^2\hat{q}_\alpha ^2\right)
+\omega_\mathrm{Vis}\boldsymbol{\lambda}_\mathrm{Vis}\hat{q}_\mathrm{Vis}\boldsymbol{\hat{\mu}}_\mathrm{el}     
     \nonumber\\
    &
   + \omega_\mathrm{IR}\boldsymbol{\lambda}_\mathrm{IR}\hat{q}_\mathrm{IR}\boldsymbol{\hat{\mu}}_\mathrm{nu}  
   +
   \frac{(
   \boldsymbol{\lambda}_\mathrm{V}\cdot
   \boldsymbol{\hat{\mu}}_{\mathrm{Tot}})^2
   +(\boldsymbol{\lambda}_\mathrm{IR}\cdot\boldsymbol{\hat{\mu}}_{\mathrm{Tot}})^2}{2}.
\end{align}
Since the optical and MIR transitions are far off-resonance, we have neglected the coupling between the electronic-dipole transition and the MIR antenna,  and similarly for the phonon mode and the optical cavity. {The final term in Eq.~(\ref{eq:PF_LW_full-expansion}) is the dipole self-energy term. In our construction, it will contain three different contributions: the electronic-, the nuclear-, and the cross nuclear-electronic dipole self energies. In general, keeping all these contributions is crucial to ensure the stability of the coupled system and the existence of a groundstate~\cite{schafer2020relevance}. However, under the two-level approximation that we apply in this work, the electronic contribution is proportional to $\hat{\sigma}_x^2 = \mathbb{I}_2$ and it thus manifests as a simple energy shift of both the ground and excited states. The electronic contribution can therefore be safely neglected in our setting.}

We now move to a second quantised notation, such that $\hat{q}_\alpha = \sqrt{\hbar/2\omega_\alpha}(\hat{a}_\alpha^\dagger + \hat{a}_\alpha)$, 
yielding the Hamiltonian,  
\begin{align}
    \hat{H} &= \hat{H}_\mathrm{mol}+ \sum_{\alpha = \mathrm{V},\mathrm{IR}}\hbar\omega_\alpha \hat{a}_\alpha^\dagger \hat{a}_\alpha
     \nonumber\\
    &
+\hbar g_\mathrm{Vis}\hat{\sigma}_x(\hat{a}_\mathrm{Vis}+\hat{a}^\dagger_\mathrm{Vis})  
+\hbar g_\mathrm{IR}(\hat{b}^\dagger + \hat{b})(\hat{a}_\mathrm{IR}+\hat{a}^\dagger_\mathrm{IR})  
\nonumber\\
   &+ \hbar\eta_c
   \hat{\sigma}_x(\hat{b}^\dagger + \hat{b})
   +\hbar\delta_c
   (\hat{b}^\dagger + \hat{b})^2.\nonumber
\end{align}
Where we have introduced the standard light-matter coupling strengths~\cite{svendsen2023molecules},
\begin{align}
    \hbar g_\mathrm{Vis} =& \sqrt{\frac{\hbar\omega_\mathrm{Vis}}{2}}\boldsymbol{\mu}_{eg}\cdot\boldsymbol{\lambda}_\mathrm{Vis},\\
    \hbar g_\mathrm{IR} =& \sqrt{\frac{\hbar\omega_\mathrm{IR}}{2}}\boldsymbol{\mu}_{N}\cdot\boldsymbol{\lambda}_\mathrm{IR}.
\end{align}
 and the counter-term energies,
\begin{align}
    \eta_c =& 2\left(\frac{g_\mathrm{Vis}^2}{\omega_\mathrm{Vis}}\frac{\mu_\mathrm{N}}{\mu_{eg}} 
    + \frac{g_\mathrm{IR}^2}{\omega_\mathrm{IR}}\frac{\mu_{eg}}{\mu_\mathrm{N}}\right),\\
    \delta_c =& \frac{g_\mathrm{IR}^2}{\omega_\mathrm{IR}}
    +\frac{g^2_\mathrm{Vis}}{\Omega_\mathrm{Vis}}\left(
        \frac{\mu_\mathrm{N}}{\mu_{eg}}\right)^2,
\end{align}
where we have introduced $\mu_{\alpha} = \vert\boldsymbol{\mu}_\alpha\vert$.
Note that the counter term {stemming from the electron-nuclear dipole self energy contribution, the term proportional to 
$\hat{\sigma}_x(\hat{b}^\dagger + \hat{b})$}, will be heavily suppressed since the 
phonon mode and optical transition are far detuned from one another. We therefore 
neglect this term. 
Furthermore, we assume that the remaining counter term is dominated by its first term, such
 that $\delta_c\approx g_\mathrm{IR}^2/\omega_\mathrm{IR}$.

Finally, since the frequency associated to the optical transition is much 
larger than any other energy scale, we can make the rotating wave approximation,
placing the visible cavity coupling in Jaynes-Cummings form,
\begin{align}
    &\hat{H} \approx \hat{H}_\mathrm{mol}+ \sum_{\alpha = \mathrm{V},\mathrm{IR}}\hbar\omega_\alpha \hat{a}_\alpha^\dagger \hat{a}_\alpha
+\hbar g_\mathrm{Vis}(\hat{\sigma}^\dagger\hat{a}_\mathrm{Vis}+\hat{\sigma}\hat{a}^\dagger_\mathrm{Vis})  
\nonumber\\
&+\hbar g_\mathrm{IR}(\hat{b}^\dagger+ \hat{b})(\hat{a}_\mathrm{IR}+\hat{a}^\dagger_\mathrm{IR})  
   +\hbar\Omega_\nu \hat{b}^\dagger \hat{b}
   +\hbar\delta_c
   (\hat{b}^\dagger + \hat{b})^2.
\end{align}

\begin{figure*}[t]
    \includegraphics[height=4.4cm]{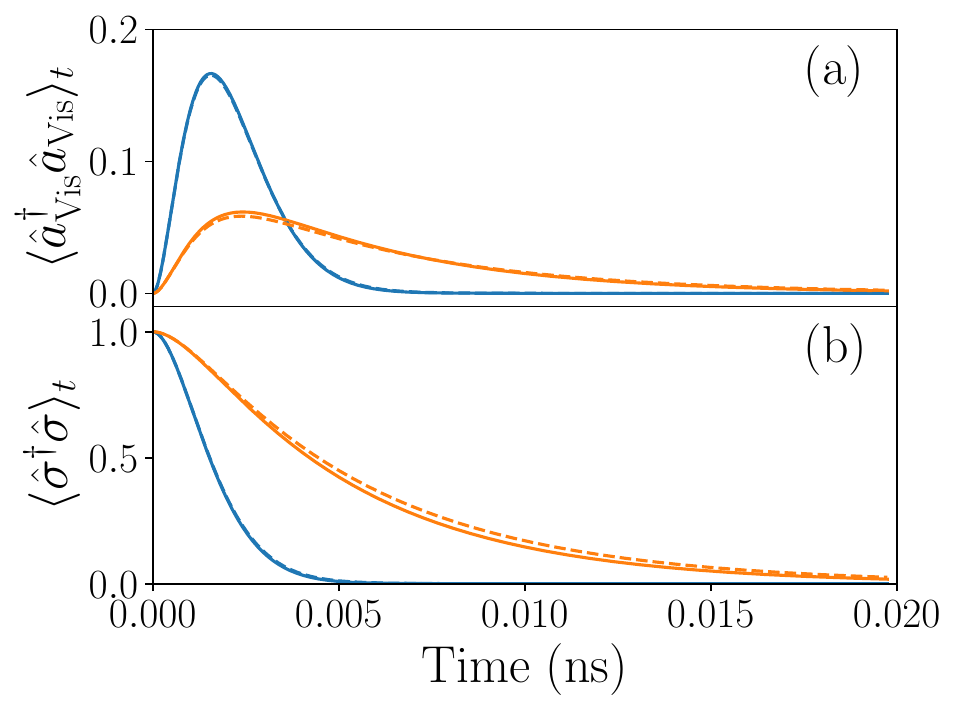}
    \includegraphics[height=4.3cm]{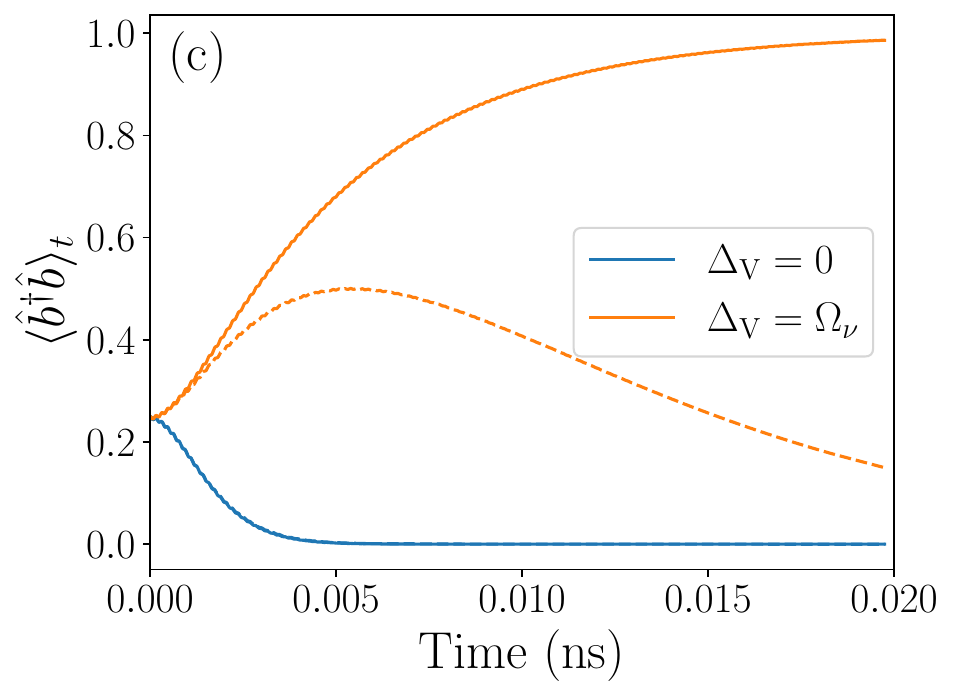}
    \includegraphics[height=4.3cm]{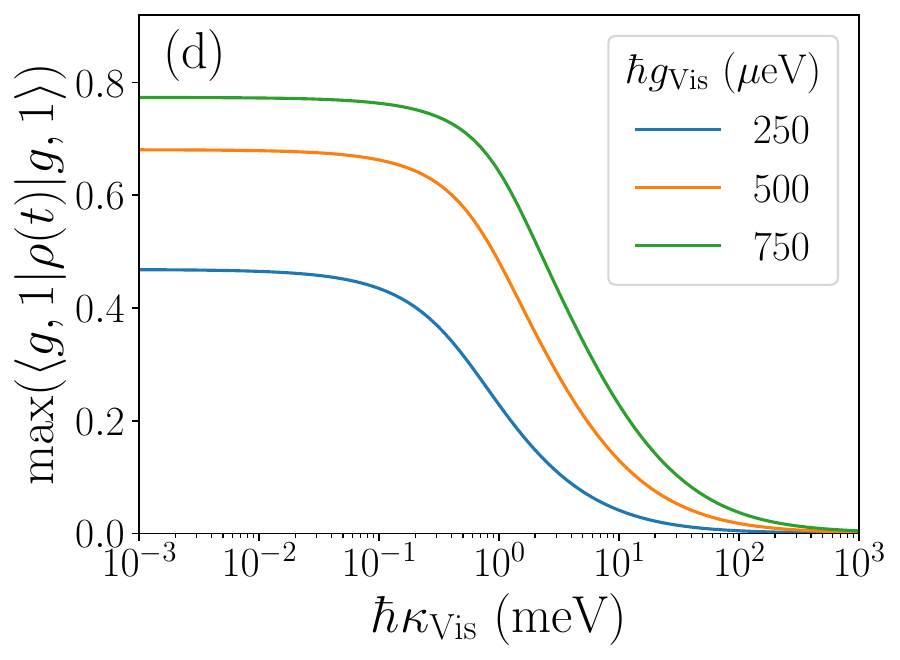}
    \caption{
    \textbf{Dynamical preparation of a single phonon state:}
    (a) and (c) show the cavity occupation and emitter dynamics respectively 
     for a cavity tuned to the  ZPL (blue) and first sideband transition (orange) 
     resonant. The dashed curves show the effect of including intrinsic losses of the 
     phonon mode. Parameters used here are $\hbar g_\mathrm{Vis}=500~\mu$eV, 
     $\hbar\kappa_\mathrm{Vis}=1.5~$meV, $\hbar\Omega_\nu = 180$~meV, 
     and $\hbar\eta_\nu=90$~meV. The intrinsic losses of the phonon mode is 
     either $\hbar\gamma=0$ (solid curves) or $\hbar\gamma=0.1$~meV.
     (c) The phonon occupation for the ZPL and first sideband enhanced emission with (dashed)
     and without (solid) intrinsic losses in the phonon mode.
     (d) The maximum transient population of the first Fock state in the ground state manifold. The intrinsic loss parameter of the phonon mode is set to $\hbar\gamma = 0.1$~meV.}
      \label{fig:dyn1}
\end{figure*}
\subsection{Master equation for the cascaded MIR source}
Now that we have a Hamiltonian for the coupled system, we wish to consider
its dynamics.
We will do this through the master equation formalism, allowing us to account for
finite lifetime of the cavity, antenna, and  phonon modes, which are assumed
to have spectral widths $\kappa_\mathrm{Vis}$, $\kappa_\mathrm{IR}$, 
and $\gamma$ respectively. Here $\kappa_\mathrm{IR}=\Gamma_\mathrm{R} + \Gamma_\mathrm{NR}$ contains both a radiative contribution $\Gamma_\mathrm{R}$ to the far field, and a non radiative contribution $\Gamma_\mathrm{NR}$ related to internal losses of the antenna (see discussion in section \ref{sec:conversion_of_phonons_to_MIR_phonons}).
Assuming all phonon and photon modes have Lorentzian profiles, 
then we may write the master equation as,
\begin{equation}\begin{split}    
    &\frac{\partial\rho(t)}{\partial t} = -\frac{i}{\hbar}\left[\hat{H},\rho(t)\right]
    +\frac{\kappa_\mathrm{Vis}}{2}\mathcal{L}_{\hat{a}_\mathrm{Vis}}[\rho(t)]\\
&    +\frac{\kappa_\mathrm{IR}}{2}\mathcal{L}_{\hat{a}_\mathrm{IR}}[\rho(t)]
    +\frac{\gamma}{2}\mathcal{L}_{\hat{b}}[\rho(t)]
\end{split}
\end{equation}
where $\rho(t)$ is the reduced density operator of the
electronic-cavity-antenna-phonon system. We have also introduced 
the Lindblad dissipator $\mathcal{L}_{\hat{O}} [\rho] = 2 \hat{O}\rho\hat{O}^\dagger - \hat{O}^\dagger\hat{O}\rho-\rho\hat{O}^\dagger\hat{O}$.
We can solve this master equation numerically to extract the 
optical properties of the composite system in both the IR and visible 
spectral regime.

\section{Cascaded quantum system for MIR photon emission}
\label{sec:cascaded}
As shown schematically in Fig.~\ref{fig:protocol}, to generate a single photon in the MIR, we first need to generate an optical 
photon, which deterministically prepares a single phonon state, 
before converting this phonon state into a MIR photon.
In this section, we shall first consider each step in the process independently 
in order to illustrate the mechanisms involved, before combining them as a 
single cascaded process. 

\subsection{Single phonon generation}

We start by considering the process by which a single phonon state is 
deterministically prepared. 
In order to illustrate this, we initially consider
a phonon mode in the absence of the MIR antenna and intrinsic losses 
(i.e. $g_\mathrm{IR}=\gamma=0$). 
We consider the quantum emitter to be initially prepared in its excited state,
with the phonon mode prepared in the ground state of the PES associated with the 
excited electronic configuration, i.e., the coherent state $\ket{S=\eta_\nu/\Omega_\nu}$, 
and with the cavity in the vacuum state.

Upon emitting a photon through the cavity, the resulting state of the phonon mode
will be dependent on the cavity parameters. 
In the abscence of the cavity, the emission behaviour is governed by the Frank Condon
principle, in which each transition is weighted by the overlap between phonon states associated to the ground and excited state PES, that is $\chi_n = 
\vert\!\braket{S}{n}\!\vert^2$, 
where $\ket{n}$ are the Fock states associated to the electronic ground state configuration.

However, an optical cavity can suppress or 
enhance optical transitions depending on the cavity parameters. 
For example, a cavity 
resonant with the ZPL ($\Delta_\mathrm{V}=0$) will enhance emission through
the ZPL while suppressing phonon sideband processes.
This is an effect commonly used by the on-demand single photon source community to enhance
an emitter's indistinguishability and efficiency~\cite{Somaschi16,Grange17,ilessmith17phonon}.
The blue curves in Fig.~\ref{fig:dyn1}(a-c) show the dynamics for the optical cavity, emitter, and phonon mode respectively with a cavity that is on resonance with the zero-phonon line in the bad-cavity limit. 
Here we see a Purcell enhanced exponential decay from the emitter, where excitation is transferred
to the cavity mode.
If we consider the phonon occupation in Fig.~\ref{fig:dyn1}(c),
we see from an initial phonon occupation given 
by the displacement of the initial coherent state, 
$\langle\hat{b}^\dagger\hat{b}\rangle_{t=0} = (\eta_\nu/\Omega_\nu)^2 = 0.25$, 
the phonon occupation drops to the ground state of the PES associated to the 
ground electronic configuration. 
Thus, by enhancing the ZPL transition, we quasi-deterministically 
prepare a phonon in its 
ground state $\ket{g,0}$ in the long-time limit.

Within this protocol, we wish to deterministically 
generate a single phonon state.
We do this by red-detuning the visible cavity to the first phonon sideband 
transition ($\Delta_\mathrm{V}=-\Omega_\nu$), selectively enhancing the 
$\ket{e,S}\rightarrow\ket{g,1}$ transition as shown schematically in Fig.~1(i), 
while suppressing other emission channels.
As shown in Fig.~2(a-b) by the orange curves, this results in a lower Purcell enhancement than the resonant cavity, 
as the dipole transition 
rate is reduced by the Frank-Condon factor $\chi_1$. 
In contrast to the $\Delta_V = 0$, the phonon population increases as a function of time,  
trending to unity in the long-time limit, indicating that the $\ket{g,1}$ state is populated.
\begin{figure*}[t]
    \includegraphics[height=4.5cm]{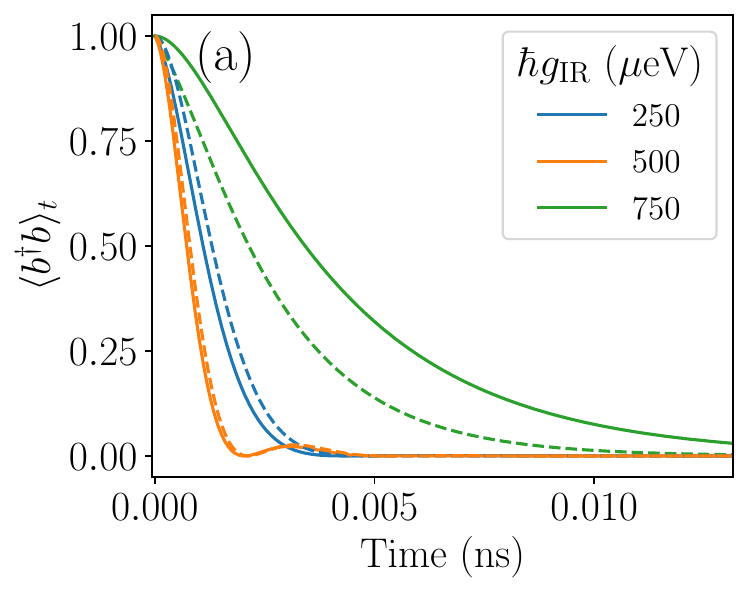}
    \includegraphics[height=4.5cm]{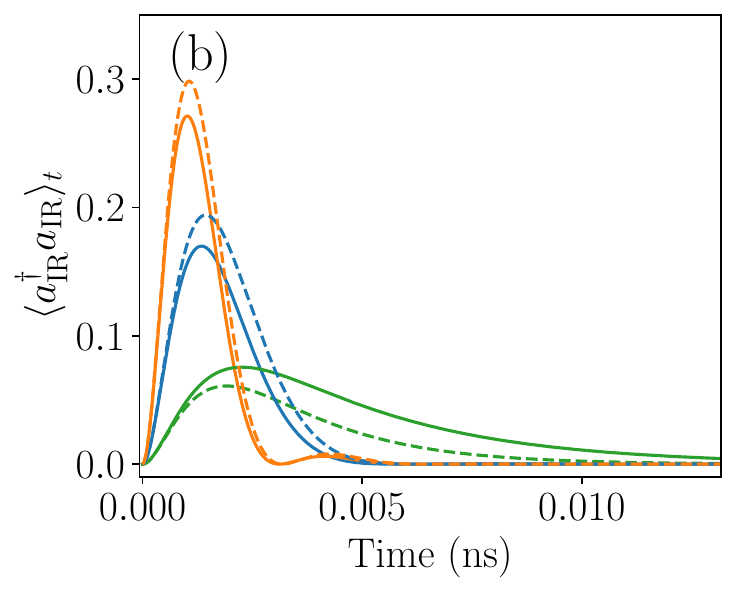}
    \includegraphics[height=4.5cm]{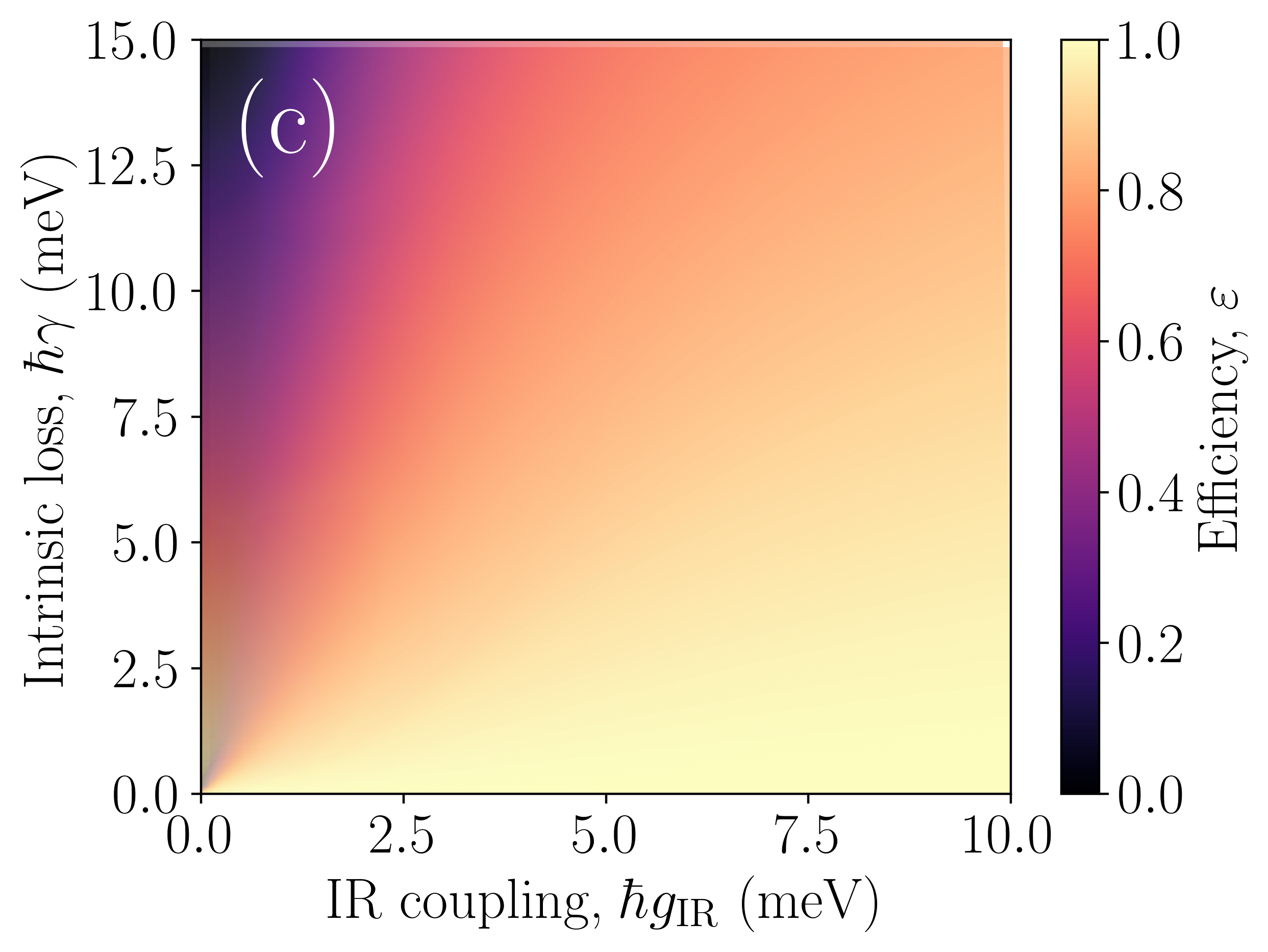}
    \caption{\textbf{The dynamics and efficiency of converting a single phonon into a MIR photon:} (a) The occupation dynamics of the phonon mode when initialised on the $\ket{1}$ 
    state with (dashed) and without (solid) intrinsic losses and an antenna loss
    rate of $\hbar\kappa_\mathrm{IR} =1.5$~meV.
    (b) The corresponding population dynamics of the MIR cavity mode. 
    (c) The conversion efficiency of a phonon to a MIR photon as defined by Eq.~\ref{eq:eff} as a function of the intrinsic loss of the phonon mode $\gamma$ and coupling strength to the MIR cavity mode $g_\mathrm{IR}$. The efficiency was calculated with a loss rate of the MIR cavity of $\kappa_\mathrm{IR}=65$~meV with  $\Gamma_\mathrm{NR}=0$.
     Other parameters are 
     $\hbar g_\mathrm{Vis} = 0.5~$meV, $\hbar\kappa_\mathrm{Vis} = 0.1$, 
    $\omega_\mathrm{IR} = \Omega_\nu$, and $\hbar\gamma=0.1$~meV. 
    All other phonon parameters are the same as in Fig.~\ref{fig:dyn1} 
     }
     \label{fig:ant_dyn}
\end{figure*}

When intrinsic loss processes are included for the phonon mode, i.e., $\gamma\neq0$, the phonon
population now has a finite lifetime, 
and thus decays to the ground-state in the long-time limit, as shown by the dashed curve in Fig.~2(c). 
However, there remains a 
significant increase in phonon occupation at early times, suggesting that $\ket{g,1}$ is still prepared. 
Fig.~2(d) shows the maximum population in the first Fock state in the ground 
state manifold (i.e. $\ket{g,1}$) as a function of cavity 
widths for a range of coupling strengths. 
We see that the maximum population monotonically increases with both increasing light-matter coupling strength ($g_\mathrm{Vis}$) and decreasing 
cavity width ($\kappa_\mathrm{Vis}$). 
It is important to note, however, that while this maximum population is not unity, this is not
a true reflection of the efficiency of the preparation scheme.
The cavity drives population through the $\ket{g,1}$ state, while intrinsic losses deplete
the phonon occupation; even in regimes where the losses are high, and thus the 
maximum phonon occupation is low, the population may still move through the $\ket{g,1}$ state.
Therefore, so long as the phonon to MIR photon conversion is faster than the intrinsic losses, then the conversion efficiency may still approach unity. We will discuss this in the following section. 

It is important to note that while we have focused on single-phonon state preparation, a similar strategy can be used to prepare multi-phonon Fock-states. By tuning the optical cavity to the $N^\mathrm{th}$-phonon sideband, one can prepare the phonon mode to be in the state $\ket{N}$ after photon emission.
With the extraction process described in the subsequent section, it is then possible to generate $\ket{N}$-photon states in the MIR.

\subsection{Converting single phonon states to MIR photons}\label{sec:conversion_of_phonons_to_MIR_phonons}

In the second part of the protocol illustrated in Fig.~1(ii), we examine the efficiency of radiating a single phonon state as a mid-infrared (MIR) photon through its interaction with an MIR antenna, following the mechanism outlined in Sec.~\ref{sec:MIRmodel}.
We start by tuning the MIR antenna to be resonant with phonon mode 
($\omega_\mathrm{IR}=\Omega_\nu$), and take the emitter and cavity modes to be in their ground states. 
The phonon mode is then initialised in the single-excitation Fock state $\ket{1}$. 
Fig.~\ref{fig:ant_dyn}(a) shows the phonon mode occupation as a function of 
time both with (solid) and without (dashed) intrinsic losses $\gamma$ for several 
different MIR-cavity parameters.
Here we see a relaxation of the phonon population to the ground state, the rate of which is Purcell enhanced by the MIR-cavity.
This behaviour is reflected in the dynamics of the MIR antenna, as shown by 
Fig.~\ref{fig:ant_dyn}(b)
where we see population accumulating in the antenna, before decaying exponentially.
As expected, the total occupation in the MIR antenna is reduced when intrinsic losses of the phonon mode are incorporated.

It is clear that the occupation of the MIR antenna is sensitive to both the cavity parameters, intrinsic loss of the phonon mode, and the non-radiative loss of the antenna.  
Therefore, it is necessary to consider the efficiency with which MIR phonons are converted to MIR photons. 
We do this in analogy to the efficiency defined in cavity QED~\cite{ilessmith17phonon}, where the efficiency is given by, 
\begin{equation}\label{eq:eff}
    \varepsilon = \frac{\Gamma_\mathrm{R} \mathcal{P}_\mathrm{MIR}}{
        \gamma\mathcal{P}_\mathrm{NR} + (\Gamma_\mathrm{R} + \Gamma_\mathrm{NR}) \mathcal{P}_\mathrm{MIR}},
\end{equation}
where we have defined the powers as,
\begin{equation}
    \mathcal{P}_\mathrm{MIR} = \int\limits_{0}^\infty 
    \langle a^\dagger_\mathrm{IR}a_\mathrm{IR}\rangle_t~\dd t ~~\text{and}~~
    \mathcal{P}_\mathrm{NR} = \int\limits_{0}^\infty 
    \langle b^\dagger b\rangle_t~\dd t.
\end{equation}
The efficiency, as defined by Eq.~\ref{eq:eff} is plotted in Fig. 4 as a function of IR coupling-strength and intrinsic losses of the phonon mode for an idealised MIR antenna with $\Gamma_\mathrm{NR}=0$. 
Here we see that there is clear competition between intrinsic losses of the phonon mode and the Purcell enhanced emission rate through the MIR antenna determined by the MIR coupling strength $\hbar{}g_\mathrm{IR}$. 
To ensure efficient conversion of the phonon to MIR, the Purcell enhanced emission rate must exceed the intrinsic losses of the phonon mode. 
For small values of $\gamma$, almost unity efficiency can by achieved with moderate values of the coupling strength $g_\mathrm{IR}$.
However, as the intrinsic loss of the phonon increases, one needs increasingly strong phonon-antenna couplings to remove the excitation via the MIR antenna.
To account for non-radiative losses of the MIR antenna, one can scale the absolute efficiency by the fraction of photons emitted through the radiative path.
We refer the reader to the supplementary information for a consideration of the impact of non-radiative antenna losses on the efficiency. 

A final consideration for the efficacy of the MIR photon generation is the purity of the resulting photons, i.e. the degree of anti-bunching of the emitted MIR photons.
We can determine this by considering the second-order 
correlation function 
of the cavity, such that,
\begin{equation}\label{eq:anti}
g^{(2)}(\tau)= N\int\limits_0^\infty 
\langle a^\dagger_\mathrm{IR}(t)a^\dagger_\mathrm{IR}(t+\tau)
a_\mathrm{IR}(t+\tau)a_\mathrm{IR}(t)\rangle
\dd t,
\end{equation}
where $N^{-1}=\mathcal{P}_\mathrm{MIR}^2$ is a normalisation constant.
The degree of bunching can be calculated from the coincidence counts at zero time delay, that is, $g^{(2)}(\tau=0)$. 
This is shown in Fig.~\ref{fig:g2}(a), which shows the MIR photons with anti-bunching several orders of magnitude below the classical limit (shown by the dashed red curve) for all parameters chosen. 
This implies that the photons emitted from the MIR cavity are highly antibunched.

\subsection{Heralded MIR photon generation}
\begin{figure*}[ht!]
\includegraphics[width=\textwidth]{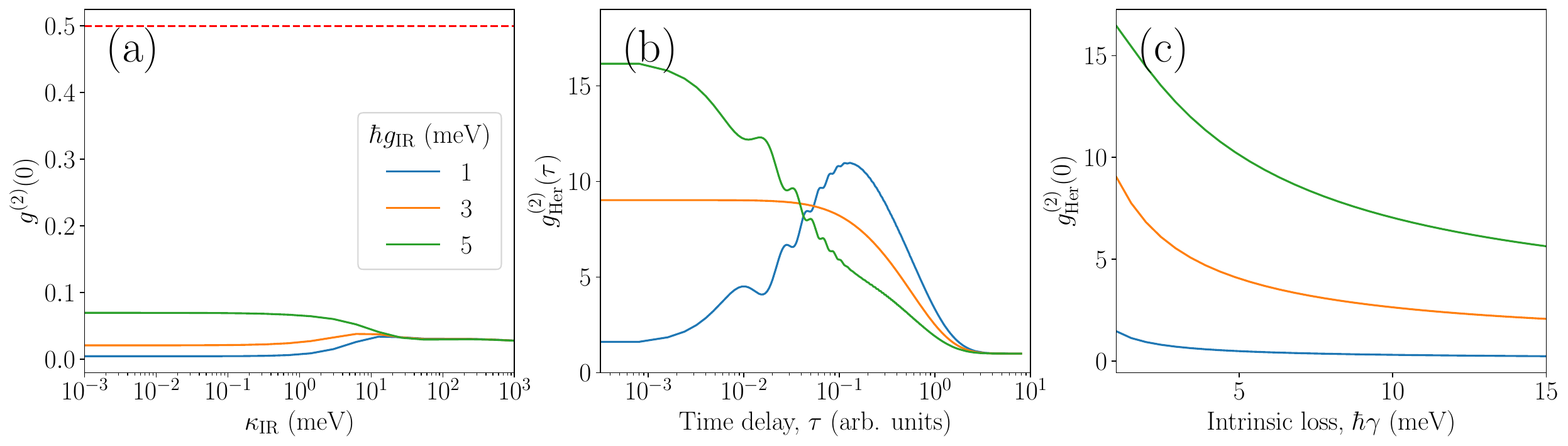}\caption{
\textbf{Correlations of photons emitted from the MIR cavity:} 
(a) A graph showing the degree of anti-bunching for the MIR photon emission calculated from Eq.~\ref{eq:anti} as a function of the MIR cavity losses and for several different values of the MIR coupling given in the legend. The dashed red curve shows the classical limit.
(b) The correlations between the heralding photon and the MIR photon emission as a function of time delay with intrinsic loss $\hbar\gamma = 1$~meV, and different values of $\hbar g_\mathrm{IR}$ matching those used in (a).
(c) Shows the change in the heralding efficiency as a function of intrinsic loss of the phonon mode.
All other parameters used match those of Fig.~3}
\label{fig:g2}
\end{figure*}

Before considering realistic material parameters and systems, we wish to confirm that the MIR photon can be heralded by the emission of a photon from the optical cavity. 
To do this, we consider the cross correlation function,
\begin{equation}
    g^{(2)}_\mathrm{Her}(\tau) = N\int\limits_{0}^{\infty} \langle a^\dagger_\mathrm{Vis}(t) 
    a_\mathrm{IR}^\dagger(t+\tau)
    a_\mathrm{IR}(t+\tau)
    a_\mathrm{Vis}(t)\rangle\dd t
\end{equation}
where $N^{-1}=\mathcal{P}_\mathrm{Vis}\mathcal{P}_\mathrm{MIR}$ is a normalisation constant.
For the MIR photon to be heralded by the visible photon $g^{(2)}_\mathrm{Her}(\tau)$ should be strongly bunched, that is, at zero time delay we expect super-poissonian statistics in the cross-correlation measurements $g^{(2)}_\mathrm{Her}(0)>1$. Fig.~\ref{fig:g2}(b) shows this behaviour for different values of the IR coupling strength $g_\mathrm{IR}$, and a fixed intrinsic loss rate of $\hbar\gamma = 1$~meV. 
Here we see distinct behaviour depending on the value of $g_{\mathrm{IR}}$: 
for the weakest coupling ($\hbar{}g_{\mathrm{IR}}=1$~meV)
there is an initial value of bunching, which becomes stronger with increasing $\tau$, suggesting a delay between detecting the heralding photon and the MIR emission.
For increasing $g_{\mathrm{IR}}$, and thus Purcell enhancement, this delay is suppressed and the initial bunching increases; this is a consequence of a faster conversion from phonon to MIR photon.

We note in passing that the increase in $g^{(2)}_\mathrm{Her}(\tau)$ at later time for weaker coupling could suggest that \(\max\limits_{\tau}\) $g^{(2)}_\mathrm{Her}(\tau)$ is a more appropriate measure of heralding efficiency. However, here we consider instantaneous heralding ($\tau = 0$) to be the experimentally relevant measure since the heralding and MIR photons should ideally fall within the same detector time bin.

The bunching of the visible and mid-IR photons is also affected by the intrinsic loss of the phonon mode, as shown by Fig.~\ref{fig:g2}(c), where we observe a decrease in the bunching $g^{(2)}_\mathrm{Her}(0)$ with increasing $\gamma$. 
This reflects a drop in the MIR photon generation efficiency, which consequently reduces the photon heralding efficiency. 
However, similar to our efficiency metric, the heralding success rate can be greatly improved by increasing the coupling strength (and thus Purcell enhancement) between the phonon mode and the MIR antenna.

\section{Realistic parameters for 
solid-state quantum emitters}
\label{sec:materials}

In this section, we will assess the material parameters of well-known quantum emitters to gauge the experimentally viable coupling strengths between polar phonon modes and an MIR antenna. The interaction between an antenna and a phonon mode is primarily influenced by two factors: the electric field strength generated by the plasmonic or photonic structure and the macroscopic polarization of the material system stemming from the polar phonon mode. Engineering efforts can enhance the electric-field strength through reduction in the mode volume. This can be achieved in nanometre-sized gapped metallic structures, as seen in plasmonic cavities \cite{baumberg2019extreme,chikkaraddy2023single,xomalis2021detecting}, or by computer-aided design in dielectric structures to significantly confine electromagnetic (EM) field modes \cite{albrechtsen2022nanometer,xiong2024experimental,choi2017self,wang2018maximizing,mignuzzi2019nanoscale}. 
Furthermore, the polarizability of a phonon mode is intrinsically tied to the chosen material platform. 

In the following, we will focus on three distinct platforms for in-depth analysis: color centers in monolayer hexagonal boron nitride (hBN), colloidal quantum dots, and organic molecules. Each platform offers unique properties and challenges in terms of phonon-antenna coupling strength, primarily due to their material-specific polarizability and phonon modes.

\subsection{Estimating the effective dipole moment of a single phonon}
For a bulk material, the polarization density induced by the excitation of phonon mode $i$ can be expressed as~\cite{resta2007theory},
\begin{align}\label{eq:pol_from_born}
    \Delta\boldsymbol{P}_i = \frac{1}{\Omega}\sum_\alpha^{N}Z_\alpha \boldsymbol{\varepsilon}_{i\alpha}.
\end{align}
Here $Z_\alpha$ is the Born effective charges of atom $\alpha$ and  $\boldsymbol{\varepsilon}_{i\alpha}$ is the displacement of atom $\alpha$ due to the phonon mode $i$. 
$\Omega$ is the unit cell volume and $N$ is the number of atoms in the unit cell. (In this work, we calculate the Born effective charges using the GPAW and ASE codes~\cite{enkovaara2010electronic,ase-paper,mortensen2024gpaw}. Details of these simulations can be found in the Supplementary Information.) Eq.~\ref{eq:pol_from_born} tells us that in order to have an induced polarization by a phonon mode, we need (i) a polar crystal (otherwise $Z_\alpha = 0\,\forall\,\alpha$), and (ii) a phonon mode which displaces the atoms such that their overall displacement results in a net movement of charge. 
In addition to the Born effective charges and the oscillation pattern, the single excitation amplitude is critical to determining the effective dipole moment of a single phonon excitation. In line with our Frank-Condon model of the phonon modes, we will extract the single-phonon displacement amplitude via the well known result for the quantum harmonic oscillator which states that,
\begin{align}
    \langle n| \hat{x}_\nu^2 |n\rangle = (2n +1)\frac{\hbar}{2\mu_R\omega},
\end{align}
where $\mu_R$ is the reduced ion mass of the lattice, $\omega$ is the characteristic frequency of oscillator, and $|n\rangle$ is the $n^\mathrm{th}$ vibrational eigenstate of the harmonic oscillator. Since we are interested in a single phonon mode, we take $n = 1$. \\

Next we consider the bounds of the effective dipole moment. The lower bound of the induced dipole moment is simply 0 Debeye (D). This is because even in a highly polar lattice, some vibrations patterns will result in zero net movement of charge. The upper bound is a more challenging to obtain because it will generally depend both on the material system and on a specific phonon mode. 

The paradigmatic polar phonon modes in bulk systems are the longitudinal optical (LO) and transverse optical (TO) phonon modes. These modes are characterized by the anion and cation of the lattice oscillating in opposite directions; along the direction of propagation (LO), and transverse to the direction of propagation (TO). In simple binary crystals, we can thus state that $\boldsymbol{\varepsilon}_\mathrm{anion} = -\boldsymbol{\varepsilon}_\mathrm{cation}$. We will use these prototypical phonon modes to set the upper bound by assuming that the maximum achievable single-phonon dipole moment is the same as the polarization of a single TO mode excitation in the primitive bulk unit cell.
 Consequently, the induced polarization within one unit cell by a single TO phonon is,
\begin{align}\label{eq:effective_dipole_moment_formula}
    \Omega|\Delta \boldsymbol{P}| = (|Z_\mathrm{cation}|+ |Z_\mathrm{anion}|)\left(\frac{3\hbar}{2\mu_R\omega}\right)^{1/2}.
\end{align}
This therefore leads to the single phonon dipole moment of $|\boldsymbol{\mu}_\mathrm{N}| = (|Z_\mathrm{cation}|+ |Z_\mathrm{anion}|)\left(\frac{3\hbar}{2\mu_R\omega}\right)^{1/2}$.



\subsection{Colour centres in hBN}
We will first discuss color centers in hBN. hBN 
is a popular host material for color centers producing single photons in the visible~\cite{tran2016quantum,fischer2021controlled,michaelis2022single,aharonovich2022quantum} which naturally hosts polar phonon modes due to the polar nature of its lattice \cite{wigger2019phonon,fischer2023combining}. 
For a phonon sideband of interest, one would calculate the Born effective charges and phonon modes of the lattice with the defect included. 
Eq.~\ref{eq:pol_from_born} would then be used to determine whether the oscillation pattern of the phonon mode responsible for the relevant sideband is such that an overall polarization change occurs.
The induced polarization is thus generally both color center and phonon mode specific. 
Because the number of defect-host material combinations are extremely large, we will refrain from considerations of specific defects in this work and instead use Eq.~(\ref{eq:effective_dipole_moment_formula}) to estimate the upper limit of the achievable dipole moments. In pristine monolayer hBN, the Boron atoms have a Born effective charge of $Z_B = 2.71e$ for displacements in plane, and the Nitrogen atoms have an effective Born charge of $Z_N = -2.71e$ \footnote{Note that $\sum_\alpha Z_\alpha = 0$ as expected}. For the frequency we will use the LO/TO energy of monolayer hBN of around 170 meV. As shown in Table~\ref{tab:comparing_platforms_real_freqs}, inserting these into Eq.~\ref{eq:effective_dipole_moment_formula} leads to an effective dipole moment of $1.04$ D.

\begin{table}[b]
\begin{tabular}{|l|l|}
\hline
Material platform               & Effective dipole moment \\ \hline
Bulk InP estimate (LO/TO)   & 1.05 D                    \\ \hline
Bulk h-BN estimate (LO/TO) & 1.04 D                    \\ \hline
Bulk CdS estimate (LO/TO) & 0.9 D                    \\ \hline
Benzene  (Pyrene freq. Ref.~\cite{weatherly23})                       & 0.11 D                \\ \hline            
\end{tabular}
\caption{\textbf{Comparing material platforms based on the bulk TO mode frequencies:} Effective dipole moment for a single phonon excitation in different material platforms with bulk TO phonon frequencies and Born effective charges calculated using Density Functional Theory.}
\label{tab:comparing_platforms_real_freqs}
\end{table}

\subsection{Colloidal quantum dots}
Colloidal quantum dots are another interesting platform for the scheme we envision. In particular, in addition to strong coupling to acoustic sidebands, the emission spectra of InP/ZnSe QDs show clear signs of coupling to the LO/TO mode of the InP crystal in the form of a sideband around 45 meV from the ZPL~\cite{brodu2018exciton,proppe2023highly}. When calculating the Born effective charges, these come out to 2.6e and -2.6e for In and P respectively. Performing the same analysis as for h-BN, we thus arrive at an effective dipole moment of 1.05 D. Here it is important to note that unlike the color centers, the QDs actually contain bulk InP. It is therefore reasonable to expect this estimate for bulk InP to be a good measure of what one would measure with a real QD.  

Giant shell CdS/Cd colloidal quantum dots display interesting quantum optical properties \cite{chen2008giant, morozov2023purifying}. These materials host also optical phonon modes featuring high dipole moments (see Table~\ref{tab:comparing_platforms_real_freqs}). Indeed, strong coupling between optically active (surface) phonon modes in CdS and metallic antenna resonant at THz frequencies has been demonstrated recently \cite{jin2018reshaping}, thereby also demonstrating the feasibility to couple efficiently nanocrystals with optical antennas at these frequencies. Calculating the Born Effective charges of CdS leads to 2.1e and -2.1e for the Cd and S atoms respectively. Following the same approach as outlined above, we find a dipole moment of 0.9 D. We note that this number is around an order of magnitude lower than the dipole moment of 17 D that one can extract from the experiment in Ref.~\cite{jin2018reshaping}. This emphasizes that more polar modes than the bulk ones can exist near surfaces (and potentially in defects such as in hBN).

\subsection{Organic molecules}
Finally, for comparison, we consider infrared active modes in organic molecules such as methylene blue or PAA where the dipole moment of infrared active mode can range from 0.1D \cite{chikkaraddy2023single} to almost 0.5~D~\cite{weatherly23}, respectively.
As a model organic molecule, we calculate the effective charges of the atoms in a benzene molecule. Our calculations show a maximum effective charge of $0.13e$ and 
$-0.13e$ for the hydrogen and carbon atoms respectively which is far lower than for the inorganic crystals. Following the same procedure as above for the inorganic crystals, we estimate a dipole moment of 0.11~D as the upper bound for the optically active phonons in benzene. This dipole moment is an order of magnitude lower than for the III-V materials and hBN.\\ 

\begin{figure}
\includegraphics[width=\columnwidth]{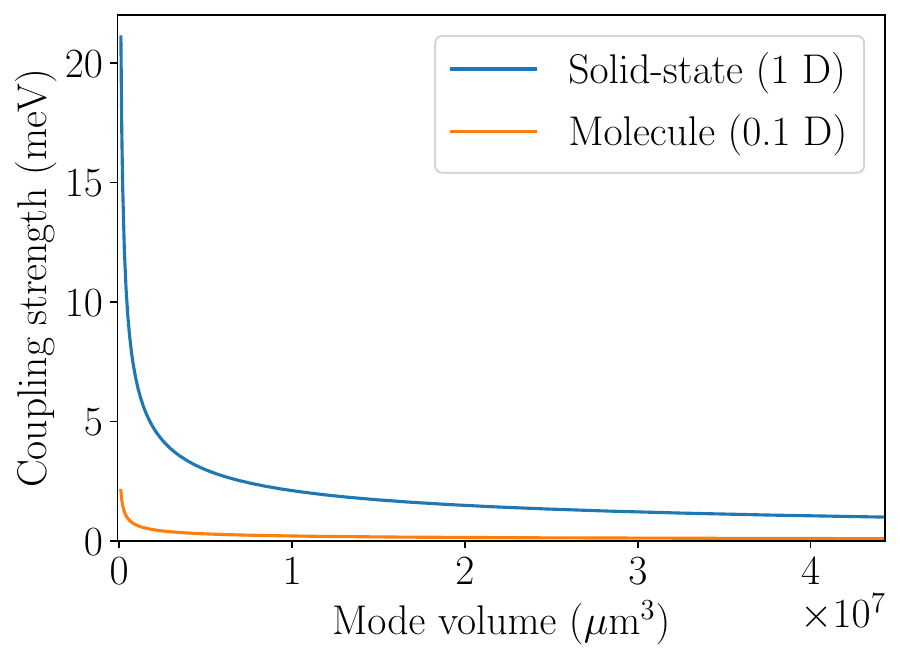}
\caption{\textbf{Antenna-phonon coupling strength as a function of mode volume:} The expected coupling strength for between a phonon modes with dipole moments characteristic of solid-state (1~D) and molecular (0.1~D) systems. For both systems, we have chosen a material permittivity of $\varepsilon=1.5$ and for a phonon mode with energy $\hbar\omega_\mathrm{IR} = 170~$meV.}
\label{fig:coupling}
\end{figure}

\subsection{Extracting antenna-phonon coupling strength}
Given the dipole moment of a material system, we can estimate the expected antenna-phonon coupling strength using Eq.~6, which assuming the dipole moment of the phonon is parallel to the field, we obtain $\hbar g_\mathrm{IR} = \sqrt{\hbar\omega_\mathrm{IR}/2}\vert\boldsymbol{\mu}_\mathrm{N}\vert\vert\boldsymbol{\lambda}_\mathrm{IR}\vert$.
This coupling strength will be dependent on the mode volume, $V_m$, of the antenna structure through the equation 
$\vert\boldsymbol{\lambda}_\mathrm{N}\vert\approx1/\sqrt{\varepsilon\varepsilon_0V_m}$, where $\varepsilon_0$ ($\varepsilon$) is the vacuum (material) permittivity and $V_m$ is the mode volume.
Fig.~\ref{fig:coupling} shows the coupling strength as a function of mode volume for phonon dipole moments characteristic of solid-state and molecular systems.
Here we see that for small mode volumes we can obtain coupling strengths necessary for efficient conversion from phonon to  MIR photon; these mode volumes are achievable with state-of-the-art plasmonic devices~\cite{Benz16,Sheinfux24} and could soon be reached experimentally with dielectric nanocavities~\cite{Babar23}.

Our analysis here suggests that III-V materials as well as highly polar 2D materials such as hBN are promising platforms for the transduction process proposed in this work. 

\section{Conclusion}

In this work we have demonstrated a unique mechanism to efficiently and quasi-deterministically generate heralded single photons at mid-infrared and even THz frequencies. 
Importantly, the parameters required for our protocol, including optical transition energies, photonic cavities, optical phonon modes, and MIR antenna coupling strengths, are all realistic and achievable with state-of-the-art materials and nanofabrication. 
{For example, coupling amplitudes between optical transitions in QE and nanocavities can reach the tens of meV regime~\cite{park2019tip,gross2018near}, and coupling strength between optical phonon modes and MIR-THz antennas of several meV have been demonstrated experimentally \cite{jin2018reshaping}.} 
Double antennas devices featuring resonances at visible and MIR frequencies have been proposed theoretically~\cite{roelli2020molecular} and demonstrated experimentally~\cite{chen2021continuous, xomalis2021detecting}. Such a design would allow emission rate enhancement of the optical transitions at the first PSB while allowing to convert single polar phonons into MIR photons sent to the far-field. We are therefore confident that our scheme is within reach experimentally using  existing technologies. 

Our scheme is versatile and may be realised with a wide range of material systems. 
In particular, polar materials such as low-dimensional hBN and 3D bulk III-V materials host optical phonon modes with dipole moments 
one order of magnitude higher than for organic molecules, thus allowing more efficient coupling to dielectric or metallic antennas.
Our scheme can also be applied to the generation of single polaritons propagating at the surface of materials~\cite{basov2016polaritons,low2017polaritons,basov2020polariton}, as long as the degrees of freedom of the polarization wave couple strongly to localized electrons or excitons in quantum emitters. 
The strong correlation between the single photon emitted in the visible and in the MIR works as an interface between the MIR-THz and the Visible-NIR regimes, thereby opening the possibility for quasi-deterministic transduction protocols and could be used in new high fidelity quantum measurements~\cite{kutas2022quantum}. 

Overall, our proposed design for new heralded quantum emitters at mid-infrared (MIR) and terahertz (THz) frequencies could enable the creation of innovative spectroscopic methods at the single photon and single phonon level. This could significantly enhance our understanding of quantum phenomena in molecular biology and in novel quantum materials where optical phonons play a significant role.

\bibliography{ref}

\begin{appendix}

\end{appendix}

\end{document}